\documentclass{aa}  
\usepackage{graphicx}
\usepackage{txfonts}
\usepackage{natbib}
\bibpunct{(}{)}{;}{a}{}{,}

\begin{document}
   \title{Millimeter dust continuum emission unveiling\\ the true mass of giant molecular clouds\\ in the Small Magellanic Cloud}
   \titlerunning{mm dust emission unveilling the true mass of SMC GMCs}

   \author{C. Bot
          \inst{1,2}
          \and
          F. Boulanger\inst{2}
          \and
          M. Rubio\inst{3}
          \and
          F. Rantakyro\inst{4}
          }

   \offprints{C. Bot}

   \institute{UMR7550, Centre de Donn\'ees Astronomiques de Strasbourg (CDS), Universit\'e Louis Pasteur,, F-67000 Strasbourg, France\\
              \email{bot@astro.u-strasbg.fr}
         \and
             Institut d'Astrophysique Spatiale, UNiversit\'e Paris-Sud, F-91405, Orsay, France\\
             \and
             Departamento de Astronomia, Universidad de Chile, Casilla 36-D, Santiago, Chile\\
             \and European Southern Observatory, Casilla 19001, Santiago 19, Chile
             }

   \date{Received ...; accepted ...}

% \abstract{}{}{}{}{} 
% 5 {} token are mandatory
 
  \abstract
   {CO observations have been so far the best way to trace molecular gas in external galaxies, but in low metallicity environments the gas mass deduced could be largely underestimated due to enhanced photodissociation of the CO molecule. Large envelopes of H$_2$ could therefore be missed by CO observations. }
   {At present, the kinematic information of CO data cubes are used to estimate virial masses and trace the total mass of the molecular clouds. Millimeter dust emission can also be used as a dense gas tracer and could unveil H$_2$ envelopes lacking CO. These different tracers must be compared in different environments.}
   {This study compares virial masses to masses deduced from millimeter emission, in two GMC samples: the local molecular clouds in our Galaxy ($10^4$--$10^5$M$\odot$), and their equivalents in the Small Magellanic Cloud (SMC), one of the nearest low metallicity dwarf galaxy.}
   { In our Galaxy, mass estimates deduced from millimeter (FIRAS) emission are consistent with masses deduced from gamma ray analysis (Grenier et al. 2005) and therefore trace the total mass of the clouds. Virial masses are systematically larger (twice on average) than mass estimates from millimeter dust emission. 
This difference decreases toward high masses and has already been reported in previous studies. This is not the case for SMC giant molecular clouds: molecular cloud masses deduced from SIMBA millimeter observations are systematically higher (twice on average for conservative values of the dust to gas ratio and dust emissivity) than the virial masses from SEST CO observations. The observed excess can not be accounted for by any plausible change of dust properties. Taking a general form for the virial theorem, we show that a magnetic field strength of  $\sim 15\mu$G in SMC clouds could provide additional support  to the clouds and explain the difference observed. }
   { Masses of SMC molecular clouds have so far been underestimated. Magnetic pressure may contribute significantly to their support.}


   \maketitle
%
%________________________________________________________________

\section{Introduction}

Molecular clouds are of particular importance as the sites of star formation in a galaxy. Most of this gas is located in large complexes known as giant molecular clouds (GMCs), but the exact amount of gas contained remains unknown since H$_2$ is quite impossible to observe directly in cold interstellar regions where molecules form and survive.

Carbon monoxyde (CO) has played a major role in GMC studies and is still the most widely used tracer of molecular gas. However, it could be present only in the densest parts of molecular clouds, leaving H$_2$ envelopes unseen. This effect has been clearly observed in Molecular Clouds of the Solar Neighbourhood \citep{GCT05} and tends to underestimate molecular cloud masses estimated from CO observations. This bias is most important in low metallicity environments where the reduced shielding causes CO to be confined to tiny dense structures \citep{ALS+04}. Most of the mass may then lie in the cloud envelopes where the gas is molecular due to H$_2$ self-shielding but CO is photodissociated \citep{Israel:2005lr,LLD+94}. Millimeter dust emission could bring to light the total mass of such clouds. 

The advent of sensitive bolometer arrays on sub-millimeter/millimeter telescopes makes long wavelength dust emission an alternative to CO to compute molecular masses.  The good correlation with CO emission in CO bright galaxies (e.g. NGC891 \citet{GZM+93}, NGC6946 \citet{Alton:2002lr}) asserts its association with molecular gas. More generally existing observations show that sub-millimeter dust emission from galaxies traces the total hydrogen column density \citep{Thomas:2002fk,Meijerink:2005qy}. It complements the H{\sc i} and CO observations to provide a complete view of cold interstellar matter.

We propose to use dust continuum emission at millimeter wavelengths to compute molecular clouds masses in two samples of GMCs: molecular clouds in the Solar neighbourhood, and counterparts in the Small Magellanic Cloud (SMC).  The SMC is a nearby irregular dwarf galaxy (D$\approx 61\mathrm{kpc}$; \citet{LS94}) known to have a low metallicity (Z$\approx$Z$_\odot/10$, \citet{Duf84}) and giant star forming regions. While the amount of molecular material observed via CO emission is low, numerous young stars are observed. Furthermore, galaxy counts behind the SMC imply extinctions much larger than expected from the column density of atomic hydrogen \citep{Leq94}. Thus the SMC is a good candidate to detect extended envelopes of H$_2$ gas  through dust observations. 

From the comparison of GMCs CO luminosity and virial mass, the SMC $X_{CO}=N(H_2)/I_{CO}$ conversion factor is estimated to be 10 times the Galactic value \citep{RLB93,MRM01}. This virial calibration of $X_{CO}$ should take into account the cloud mass in the CO-poor H$_2$ gas, but dust observations question this view. \citet{LIS+02}  inferred a $X_{CO}$ conversion factor 25 to 30 times higher than the Galactic value by comparing sub-millimeter dust and CO emission from the low metallicity dwarf galaxy NGC1569. \citet{RBR+04} reported the detection of the millimeter dust emission from SMCB1\#1 with SIMBA/SEST. They converted the flux into a cloud mass estimate that is found to be one order of magnitude larger than its CO virial mass. \citet{Leroy:2006fk} analysed Spitzer far infrared maps in the SMC, in conjunction  with H{\sc i} and CO observations. Comparing the 160$\mu$m map to NANTEN CO observations they estimate $X_{CO}$ to be 60 times the Galactic value. This present study follows and continues the work of \citet{RBR+04} by analysing a larger number of molecular clouds in the SMC and by analysing these clouds together with a reference sample that we build in our Galaxy. As such it is the first comprehensive comparison of two giant molecular cloud mass estimates at different metallicities: gas masses from dust millimeter emission and virial masses. Are the dust and virial mass estimates in agreement? If they are discrepant, how can the difference be explained?
 
After an overview of the data used in this study (Sect. \ref{sec:data}), molecular gas masses from dust millimeter emission are computed for a sample of Local GMCs and for detected SMC GMCs (Sect. \ref{sec:mm2molmass}). A comparison to virial mass estimates  is performed in Sect. \ref{sec:comparmass} and the differences are interpreted in Sect. \ref{sec:result}

\section{The data}\label{sec:data}

This study follows the detection of several molecular clouds at 1.2~mm in the SMC with the SIMBA bolometer. We present these new observations in details, along with previously published data.

\subsection{SEST observations in the SMC}

 \subsubsection{CO observations}\label{sec:cosmc}
 
 The CO observations of the SMC clouds were observed as part of the ESO/SEST Key Programme on CO in the Magellanic Clouds \citep{IJL93,RLB93,RLB+93,RLB+96,IJR+03}. The motivation for undertaking these observations in the transitions J=(1-0) and J=(2-1) of the CO molecule, and the reduction treatments are described in these papers. We just recall here the spatial resolution: the beam has a FWHM of 43\arcsec and 22\arcsec at the CO(1-0) and the CO(2-1) frequencies respectively. Since CO(2-1) line data has better resolution, gas masses computed for the CO(2-1) emitting regions will be favored when CO(2-1) observations were performed. In the following, we will use the spatial (radius) and kinematic (velocity dispersion) characteristics of the clouds that were computed in the litterature.

\subsubsection{SIMBA millimeter continuum observations}

The SIMBA bolometer was mounted at the Swedish ESO Submillimeter Telescope (SEST). It operated at a central frequency of 250 GHz (1.2~mm) with a FWHM bandwidth of 90 GHz, and an angular resolution of $24\arcsec$ on the sky. The individual maps were produced using the fast scanning mode. The elimination of the correlated sky noise, adding of the individual maps, and the photometry were done using the MOPSI package\footnote{MOPSI is a data reduction software package for IR and radio data developped by R. Zylka, IRAM, Grenoble, France}. The $\tau_{zenith}$ was determined performing skydips on average 2 hours during the day, and every 3 hours during the night. The flux determination was done for each individual observing run by taking maps of Uranus. The data reduction was performed as described in the SIMBA User manual\footnote{http://puppis.ls.eso.org/staff/simba/manual/simba/}. The flux measurements in the final coadded map has an uncertainty of 15\%.

\begin{figure*}
\includegraphics[width=0.5\textwidth]{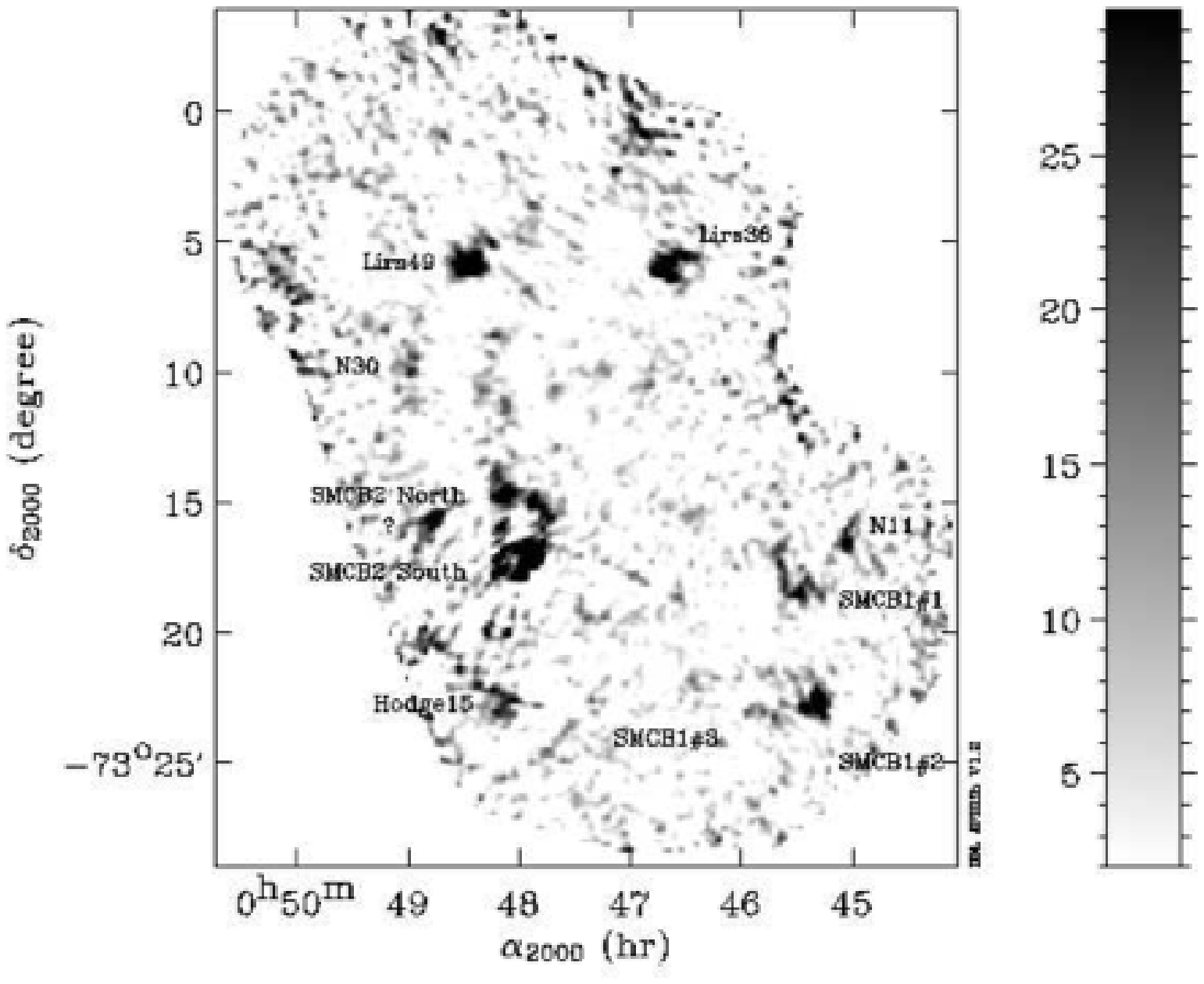}\hfill\includegraphics[width=0.5\textwidth]{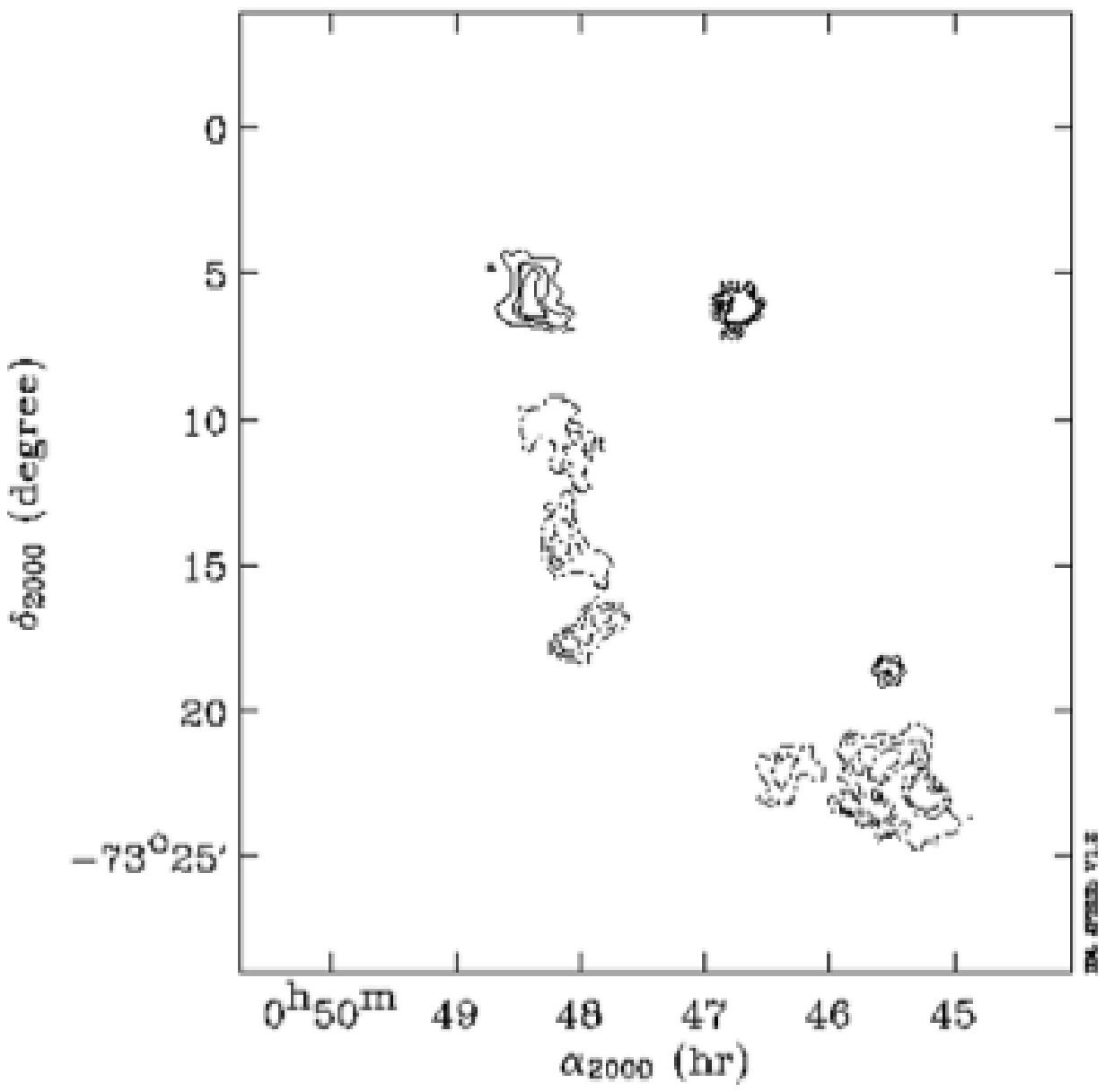}
\caption{SIMBA mosaic of the south of the SMC (left panel). Names of the star forming regions studied in this paper are labeled closeby. On the right panel, CO contours of the studied regions are displayed on the same grid (with plain lines when CO(2-1) is available, with mixed lines for CO(1-0) data). \label{fig1}}
\end{figure*}

The millimeter dust emission in the SMC is faint and the regions unambigously detected are giant molecular clouds (GMCs) (see Fig.\ref{fig1}). Most of them have been observed in CO and are named according to \citet{RLB93, RLB+93} or \citet{Hen56}. SIMBA fluxes for each region are displayed in Table \ref{tbl-1}. They are measured for a SIMBA beam convolved to $27\arcsec$ and within the $25$~mJy/beam contour, after subtraction of the flux measured in an empty zone.

\begin{table}
\begin{minipage}[t]{\columnwidth}
\centering
\caption{Observed characteristics of the detected GMCs in the SMC.\label{tbl-1}}
\begin{tabular}{llllll}
name & Area &  $S_{1.2mm}$\footnote{corrected from the contribution of free-free emission and CO(2-1) line}\\
 & pc$^2$ & mJy  \\
\hline
LIRS49 & 357  & $181\pm 55$    \\
LIRS36 & 385 & $184\pm 48$    \\
SMCB1\#1 & 108 & $47\pm 7$    \\
SMCB1\#2 & 260 & $119\pm 51$  \\
SMCB1\#3 & 38 & $5.8\pm 0.7$ \\
Hodge15 & 103 & $40.4\pm 2.5$ \\
SMCB2 S & 695 & $398\pm 158$ \\
SMCB2 N & 314 & $132\pm 46$  \\
N11     & 87 & $34.5\pm 5$  \\
\end{tabular}
\end{minipage}
\end{table}

\paragraph{Detected sources:}

Detected GMCs are often associated with well-known H {\sc ii} regions and young clusters and all but LIRS49 correspond to absorption patches \citep{Hod74}. We review here some of their properties:
\begin{itemize}
\item \emph{LIRS49} and \emph{ LIRS36} are two of the most prominent SMC sources in the far-infrared \citep{SI89} and a large variety of molecular species have been observed in these directions \citep{CHM+98,HJO98}. LIRS49 may be a superposition of two H {\sc ii} regions belonging to different kinematical complexes along the line of sight \citep{RLG94} or it could be associated with a bubble or a loop \citep{MeAz93}.

\item \emph{SMCB1\#1} and \emph{Hodge15} are two quiescent molecular clouds (not associated with H$\alpha$). The SIMBA millimeter observations of SMCB1\#1 were used by \citet{RBR+04} to deduce a cloud mass.

\item \emph{SMCB1\#2} and \emph{SMCB1\#3} are parts of the SMC-B1 region of \citet{RLB+93} and are associated with large H {\sc ii} regions. 

\item The SMCB2 region contains several molecular clouds observed by \citet{RLB93}. 
Since these clouds are difficult to separate in the continuum dust emission which has no velocity information, they were grouped into two complexes: \emph{SMCB2 North} (corresponds to SMCB2\#3 and \#4 of \citet{RLB93}) and \emph{SMCB2 South} (which corresponds to SMCB2\#1, \#2 and \#6). The SMCB2\#5 region of \citet{RLB93} is not detected at 1.2mm with SIMBA.

\item Millimeter emission is also detected toward the \emph{N30} and \emph{N11} \citep{Hen56} H {\sc ii} regions, but these regions have not yet been mapped in CO, so no mass comparison can be done for these objects. For N30, the SIMBA flux can be accounted for entirely by free-free emission.

\item The source identified by the question mark in Fig. \ref{fig1} (between SMCB2 North and South) does not correspond to any presently known source. It is clearly detected at the 6$\sigma$ level  but there is no associated infrared or radio component. It should be noted however that it lies on an extended component seen at 70$\mu$m with Spitzer observations and it is close to the DEM S43 H {\sc ii} region. Further observations are needed to understand the nature of this source.
\end{itemize}

The observed regions correspond to different evolutionary stages of molecular clouds -- from quiescent   clouds to evolved massive star forming regions--, enabling us to compare CO and dust mass estimates in different physical conditions. 
 
 \paragraph{Origin of millimeter continuum emission:}
 
 The emission measured at 1.2~mm consists of several components: thermal dust emission, free-free radiation and $^{12}$CO(2-1) line emission. Since our interest is only in cold dust emission, we estimate the other contributions that have to be removed.

\smallskip
\emph{CO line emission:} The $^{12}$CO(2-1) line contribution to the SIMBA surface brightness can be estimated as follows:
\begin{equation}
F_{line}=\frac{2k\nu^3c^{-3}}{\Delta \nu_{bol}} ~\Omega ~I_{CO(2-1)},
\end{equation}
where $I_{CO(2-1)}$ is the intensity of the $^{12}$CO(2-1) line in K.km.s$^{-1}$
, $\Delta \nu_{bol}$ is the bandwidth of the SIMBA bolometer (90 GHz), $\Omega$ is the region of integration, $k$ is the Boltzmann constant and $\nu=230$~GHz.
For regions where CO(2-1) \citep{RLB+93,RLB93,RLB+96,IJR+03} has been observed, the total flux density due to line contributions was evaluated (see Table \ref{tbl-1}) and was found to be negligible.

\smallskip

\emph{free-free emission:} The contribution of the continuum emission due to thermal electrons  can be estimated with radio continuum data, assuming pure free-free emission at high frequencies and extrapolating this frequency dependence to 1.2~mm (250 GHz) \footnote{For electron temperatures typical of H{\sc ii} regions ($\sim 10^4$K), the free-free emission scales as $(0.95-0.16 \ln \nu)$ where $\nu$ is in GHz \citep{Reynolds:1992lr}}.  We also assume that radio emission is optically thin at long frequencies. 
We used the ATCA continuum catalog of the SMC radio sources \citep{PFR+04} at 1.42, 2.37, 4.8 and 8.64~GHz.  In both SIMBA and ATCA observations, extended emission on scales larger than a given threshold is filtered out. For SIMBA, extended emission on scales larger than $\sim 2'$ is filtered out when subtracting the sky emission. The interferometric ATCA observations are not corrected for missing short spacing and do not measure emission on spatial scales larger than about half the size of the primary lobe. At 8.64 GHz, the two angular scales are comparable. 
Radio emission comes both from free-free and synchrotron emission but at these high frequencies it is dominated by free-free.
We estimate the free-free contribution by extrapolating the 8.64~GHz radio flux when available, since it is the less synchrotron biased frequency of this catalog and also the resolution is the closest to the SIMBA one. When no 8.64~GHz flux is relevant, we take the longest frequency available and follow the same procedure: for LIRS49, the free-free contribution is extrapolated from the 2.37~GHz flux; for SMCB2 North and South region it is obtained from the 4.8~GHz fluxes. It should be kept in mind in these cases that the free-free contribution could be overestimated, a bias that would reinforce the result (see section \ref{sec:result}).

\smallskip

\emph{dust emission associated with neutral atomic gas H{\sc i}}. In the SMC, the resolution of H{\sc i} observations does not allow an estimate of the contribution of the SIMBA flux that is associated with neutral gas. However, most of this emission is spatially extended and filtered out in the subtraction of the sky emission.

For each source, dust emission is the major component of the 1.2mm flux. In the following, we shall refer to the SIMBA flux tabulated in Table \ref{tbl-1} as the flux measured at 1.2~mm subtracted from the gas contribution estimates.
 
\subsection{FIRAS and CO data in our Galaxy}

To complement the analysis of SMC clouds,  we use FIRAS spectra and the CO (1-0) sky survey from \citet{DHT01} for a sample of Local GMCs defined in rectangular $(l,b)$ boxes\footnote{The box were defined on the millimeter emission map}: Orion, Taurus, Ophiuchus, Cepheus, Chamaeleon, Lupus and Perseus. The $7^o$ FIRAS beam for these clouds at a few hundreds of parsecs and the SIMBA beam at the distance of the SMC correspond to similar linear spatial resolutions. 

\subsubsection{Millimeter fluxes}

The 1.2mm fluxes were derived by fitting the FIRAS spectra at $\lambda > 500\mu$m with a black body multiplied by a power law emissivity. The temperatures in Table \ref{tbl-2} were derived from a single temperature fit of the full FIRAS spectra down to $150\mu$m and an emissivity index of 2.
As for the SIMBA observations, the 1.2mm emission observed toward the Galactic molecular clouds originate from different components other than cold dust emission. We try to estimate their contribution to the 1.2mm flux.

\begin{table*}
\begin{minipage}[t]{\columnwidth}
\centering
\caption{Observed characteristics of the Galactic GMCs studied. \label{tbl-2}}
\begin{tabular}{lllllll}
name & $l_{min}$,$l_{max}$/$b_{min}$,$b_{max}$ & distance &  $S_{1.2mm}$\footnote{corrected from the contribution of dust associated with H {\sc i}} & $T_{dust}$ & $R$ & $\Delta V$\\
 &   & pc & mJy & K & pc & km.s$^{-1}$ \\
\hline
Orion a & 204.6 , 216.4/ -23.9 , -13.6 & 450 &  $11000\pm 3000$ & 19.1 &  28.9 & 8.0 \\
Orion b & 202.1, 205.9/  -14.4 , -11.1& 450 &  $1700\pm 300$ & 19.1 &11.6 & 3.6\\
Taurus & 170.1,  188.9 /  -19.9 , -10.1 & 140 & $14000\pm 4000$ & 16.2 &13.5 & 3.8\\
Ophiucus & 347.1, 359.9 /  11.6 , 24.9& 130 &  $15000\pm 4000$ & 19.6 &11.5 & 3.8\\
Cepheus & 102.1, 116.9 /  12.1 , 20.9& 300 &  $9700\pm 4000$ & 16.1 & 22.5 & 6.8\\
Chamaeleon & 292.1, 305.9 / -19.9 , -10.6 & 160 &  $6100\pm 3000$ & 14.8 & 6.2 & 4.6\\
Lupus & 330.1, 344.9 / 10.1 , 19.9 & 140 &  $7000\pm 4000$ & 17.4 & 5.2 & 5.1\\
Perseus &156.6, 161.9 / -22.9 , -15.1 & 320  & $4700\pm 1500$ &  17.1 & 14.6 & 6.9\\
\end{tabular}
\end{minipage}
\end{table*}

\emph{The CO(2-1) line emission} contribution is difficult to estimate without CO(2-1) data on hand. However, since brightness temperatures are the same in CO J=2-1 and J=1-0 lines, the \citet{DHT01}  survey can be used to estimate this contribution. Using a $I_{CO(J=2-1)}/I_{CO (J=1-0)}=0.6$ for the Solar Neighborhood \citep{Hasegawa:1996lr}, this contribution is found to be negligeable.

\emph{Free-free emission:} extinction corrected-H$\alpha$ emission \citep{Dickinson:2003lr} is used to estimate the free-free emission contribution to the FIRAS fluxes. This is found to be very low ($<3\%$) and is therefore negligeable.

\emph{Dust emission associated with neutral atomic gas}. Unlike the SIMBA observations, the FIRAS data measures the total Galactic emission. Part of this emission comes from dust associated with the diffuse medium. We used the Leiden/Argentine/Bonn (LAB) H{\sc i} survey of \citet{Kalberla:2005ds} to estimate the proportion of millimeter emission associated with neutral gas. At high galactic latitude, the dust millimeter emission is well correlated with H{\sc i} column densities \citep{BAB+96}. The dust emission per H atom is combined with the H{\sc i} column densities measured in the cloud areas to compute and subtract the 1.2mm flux associated with neutral gas. Their contribution amounts to about 40\% of the total FIRAS flux at 1.2mm. The 1.2~mm fluxes obtained are listed in Table \ref{tbl-2} for each cloud. To estimate the uncertainties associated with the dust emission removal, we defined for each cloud a reference area of similar size on the symetric position within the Galaxy (same longitude, opposite latitude) and subtract the flux observed in the reference regions from the one observed in the molecular clouds. The differences observed are similar to the uncertainties on the data quoted in Table \ref{tbl-2}

\subsubsection{CO emission\label{sec:cogal}}

We use the CO(1-0) survey of \citet{DHT01} of the Galaxy to compute the size and velocity dispersion of each cloud over the same areas as the millimeter emission observations. To be consistent with CO information in the SMC, we chose to compute the size and the velocity dispersion of each cloud in a similar way (e.g no correction for velocity gradients). For each cloud, we fit each spectrum above a certain level and sum the gaussian fits to deduce a total cloud spectrum. The velocity dispersion of the cloud is determined by fitting a gaussian to the resulting spectrum. This method is close to what was used for the SMC except that some SMC clouds were divided into substructures when independent clouds were separated in spatial-velocity space. We checked that this difference in method for some clouds does not induce a significant bias in the cloud parameters. The radius and velocity dispersions are listed in Table \ref{tbl-2} for each cloud. 
Because there is insufficient spatial resolution to correct for velocity gradients in the SMC clouds, we chose not to correct for them in the Galactic sample either. We are careful to apply the same method to deduce virial masses in both data sets.

%%%%%%%%%%%%%%%%
%  MOLECULAR CLD MASS  %
%%%%%%%%%%%%%%%%

\section{Estimates of molecular cloud masses}

\subsection{Molecular gas mass from millimeter continuum emission}\label{sec:mm2molmass}

Dust millimeter continuum emission is associated with molecular gas.  In this section, we compute molecular cloud masses from the dust emission at 1.2mm. 

The far-infrared/millimeter emission from dust can be expressed as:
\begin{equation}
I_{\lambda}=N_H \epsilon_H(\lambda)B_{\lambda}(T_{dust})
\label{eq:emimm}\end{equation}
where $N_H$ is the hydrogen column density of the cloud, $\epsilon_H(\lambda)$ is the emissivity of dust per hydrogen atom at the wavelength $\lambda$ and $T_{dust}$ is the dust temperature. The emissivity per hydrogen atom can be expressed in terms of the absorption coefficient per unit dust mass $\kappa(\lambda)$:
\begin{equation}
\epsilon_H(\lambda)=\kappa(\lambda) x_d \mu m_H
\end{equation}
where $x_d$ is the dust-to-gas mass ratio, and $\mu m_H$ is the gas weight per H, taking into account the contribution of He. This emissivity may be determined empirically from the correlation between dust and gas observations, independently of any dust model (e.g \citet{BAB+96} for the diffuse interstellar medium in the Solar Neighborhood).

While dust temperatures for the Galactic clouds have been deduced from the FIRAS spectra fitting , this information is not available for the clouds in the SMC.  Since we wish to apply the same method for both samples, we will fix the dust temperature for both samples. Within each cloud, dust is present over a range of temperatures. The fit of FIRAS spectra with a single temperature gives higher weight to the warmer dust present at the cloud surface or near embedded stars. But in a molecular cloud, the bulk of the mass is shielded from UV radiation. The dust temperature in the shielded parts of the clouds is observed to be smaller than the values in Table \ref{tbl-2}. For example, based on submillimeter observations, \citet{SAB+03} report a dust temperature of 12~K for a Taurus filament. To conservatively estimate cloud masses, we adopt a dust temperature of 15~K (the minimum value of the FIRAS temperatures listed in Table \ref{tbl-2}) to derive Galactic and SMC cloud masses from their millimeter dust fluxes.

We assume that the dust grain opacity per unit mass is the same for the SMC and Milky Way molecular clouds. Therefore, the emissivity per hydrogen atom scales as the dust-to-gas mass ratio. Knowing the SMC dust-to-gas mass ratio and the Galactic dust emissivity per hydrogen atom in the SMC, the millimeter emission can therefore be translated into a gas amount. 
The Galactic emissivity per hydrogen atom is adopted for dust associated with molecular gas (see Appendix \ref{append1}): $\epsilon_H^{ref}(1.2mm,H_2)=(1.2\pm 0.2) \times 10^{-26} at^{-1}.cm^{2}$.
For the SMC, this emissivity will be reduced by one sixth to account for the difference in dust-to-gas ratio with respect to the Solar Neighbourhood (see Appendix \ref{append2}). The values of dust emissivity per hydrogen atom we use, correspond to absorption coefficient per unit mass of $\kappa=0.72$cm$^2$.g$^{-1}$,  3 times higher than the standard value from \citet{Draine:1984lr}. Furthermore, the dust-to-gas ratio in the SMC could be as much as 5 times smaller \citep{BBL+04}. The mass estimates we deduce from millimeter emission in the SMC are therefore upper limits. These choices were made to emphasize the robustness of the results of this study, but the range of parameters available for the dust emissivity and dust-to-gas ratio is assessed throughout the paper (see Sec. \ref{sec:result} and Fig. \ref{fig3}).

We can thus translate the measured millimeter fluxes of the clouds into their corresponding hydrogen column densities and molecular cloud masses. The results for the Solar Neighborhood and the SMC clouds are presented in tables \ref{tbl-3} and \ref{tbl-4} respectively. Note that we used an emissivity per unit dust mass and a dust-to-gas ratio that are 1.8 and 1.7 times higher, respectively, than the one used by \citet{RBR+04}. 

\begin{table}
\begin{minipage}[t]{\columnwidth}
\centering
\caption{Different mass estimates of the Galactic GMCs studied.\label{tbl-3}}
\begin{tabular}{lllll}
name & $M_H^{mm}$ & $M_H^{vir}$ & $f=\frac{M_H^{mm}}{M_H^{vir}}$\\
& $10^4$M$_\odot$ &$10^4$M$_\odot$  & \\
\hline
 Orion\footnote{the mass are obtained by adding the masses of Orion a and Orion b} & $14\pm3$ & $27\pm 2$ &   $0.5\pm 0.2$\\
 Taurus & $1.3\pm0.4$& $2.6\pm 0.1$ & $0.5\pm 0.2$\\
 Ophiucus & $1.3\pm0.3$& $2.3\pm 0.1$ & $0.6\pm 0.2$\\
Cepheus & $4.3\pm1.8$& $14.2\pm 0.8$ & $0.3\pm 0.2$\\
Chamaeleon & $0.8\pm0.4$& $1.7\pm 0.1$ & $0.5\pm 0.3$\\
Lupus & $0.7\pm 0.4$ & $1.8\pm 0.1$ & $0.4\pm 0.3$\\
Perseus & $2.4\pm0.7$& $9.3\pm 0.4$ & $0.3\pm 0.1$\\
\end{tabular}
\end{minipage}
\end{table}

\begin{table}
\centering
\caption{Different mass estimates of the SMC GMCs studied with the associated uncertainties. Notes: (1) computed from R and $\Delta V$ using CO (1-0) data; (2) computed from R and $\Delta V$ using CO (2-1) data \label{tbl-4}}
\begin{tabular}{llll}
name & $M_H^{mm}$ & $M_H^{vir}$ & $f=\frac{M_H^{mm}}{M_H^{vir}}$\\
& $10^4$M$_\odot$ & $10^4$M$_\odot$ & \\ 
\hline
 LIRS49 &  $19\pm 6$&  $8.9\pm 0.5^{(2)}$  & $2.2\pm0.7$\\
 LIRS36 & $19\pm 5$& $14.0\pm 0.5^{(2)}$ &$1.4\pm0.4$\\
 SMCB1\#1 & $5.1\pm 0.8$ &  $2.5\pm 0.2^{(2)}$ &$2.0\pm0.3$\\
SMCB1\#2 & $13\pm6$&  $7.3\pm 1.4^{(1)}$ &$1.8\pm0.8$\\
Hodge15 & $4.3\pm0.3$&  $4.9\pm 0.4^{(2)}$ & $0.9\pm0.1$\\
SMCB2 S & $43\pm17$&  $14.3\pm 2.3 ^{(1)}$ & $3.1\pm 1.2$\\
SMCB2 N & $14\pm 5$&  $9.8\pm 1.9^{(1)}$ & $1.5\pm0.5$\\
\end{tabular}
\end{table}

\subsection{Comparison to virial masses}\label{sec:comparmass}

In both data sets, the kinematic and spatial information provided by the CO data cube enables the determination of virial masses. For this determination, we use the simplified formulation for a $1/r$ density law from \citet{MRW88}:
\begin{equation}
M_{vir}(M_\odot)=190\Delta V^2(km.s^{-1})R(pc)=\frac{9\sigma^2 R}{2G}
\end{equation}
where $\Delta V$ is the observed velocity dispersion from the CO linewidth (FWHM) and $R$ is the radius of the cloud.
The virial masses for both cloud samples are computed using the radius and velocity dispersion of the clouds. For the SMC, these parameters were taken from the litterature (c.f. Sec. \ref{sec:cosmc}). For the Galactic clouds, the radius and velocity dispersion were computed as described in Sec. \ref{sec:cogal}, in order to be consistent with the SMC values. The results for Galactic and SMC clouds are presented in tables \ref{tbl-3} and \ref{tbl-4} respectively.

\citet{Rosolowsky:2006lr} have shown that spatial resolution and sensitivity differences in molecular-line data cubes can significantly affect the measured sizes and line widths that are used to deduce virial masses of giant molecular clouds. In the present study, the linear spatial resolution of the SEST CO(2-1) data in the SMC and of the CO(1-0) data from \citet{DHT01} in the solar neighbourhood are similar. Bias arising from low resolution should therefore affect the SMC and Galactic sample in the same way. This offset may be more problematic when no CO(2-1) data was available for some clouds in the SMC and CO(1-0) data with larger resolution ($\sim 13$pc) was used instead. However, the results we obtain in these cases are not significantly different that the results obtained with higher resolution CO data so the uncertainties induced do not affect the results of this paper.

Virial masses for the clouds of the Galactic sample have been published previously \citep{de-Geus:1990lr,Ungerechts:1987fk,Boulanger:1998qy,Grenier:1989uq}. However, the characteristics of the clouds were computed for substructures and the masses of the substructures were then added. We have compared our virial mass estimates with those of the litterature for the same clouds. We observe minor differences that reflect the different definitions in the extent of the clouds and the difference in the method used (velocity gradients, substructure divisions).

\section{Results\label{sec:result}}

Tables \ref{tbl-3} and \ref{tbl-4} give mass estimates for the clouds in our Galaxy and in the SMC, computed from millimeter observations of dust emission and from virial estimates. These two estimates of the gas mass of a cloud are compared through the ratio of these masses, $f=\frac{M_{1.2mm}}{M_{vir}}$.

\begin{figure}
\includegraphics[width=0.5\textwidth]{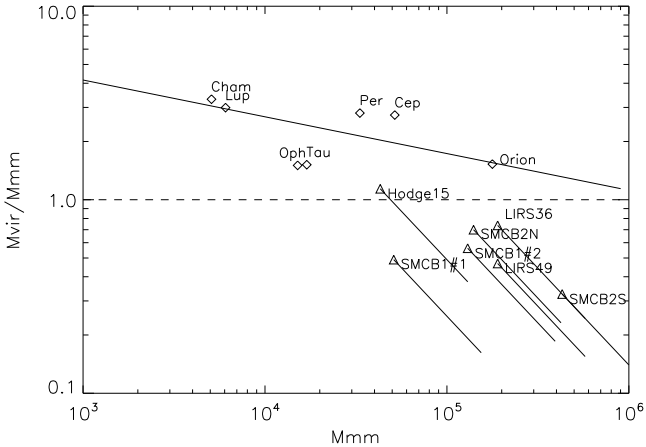}
\caption{Evolution of the ratio of the virial mass and the mass deduced from millimeter emission with respect to the mass deduced from the millimeter dust emission in solar masses $M_\odot$.The Galactic clouds are represented by rhombs while the SMC clouds are triangles. The law of  \citet{SRB+87} is also plotted (the solid line), representing the evolution of a larger sample of Galactic clouds. For the SMC clouds, the uncertainties on masses deduced from millimeter emission (regarding to the assumptions on the dust to gas ratio and the dust emissivity) are represented by a straigth line for each cloud. \label{fig3}}
\end{figure}

We observe a systematic difference between the two mass estimates:  in our Galaxy, the molecular cloud masses deduced from millimeter emission are systematically lower than the virial masses obtained from CO observations, while in the SMC the mass deduced from millimeter emission are systematically larger. This difference is best seen in Fig.\ref{fig3}.

\subsection{In the Galaxy}

The comparison of the virial mass and the gas mass deduced from millimeter emission in the local Galactic clouds shows that the virial mass is always larger than the mass from millimeter emission, with a mean $f$ factor of 0.5.

\citet{GCT05} analysed $\gamma$ ray emission in Giant molecular clouds of the Solar Neighborhood and found that a significant part of the mass of large molecular clouds is not traced by CO emission. We compared our mass estimates to their masses of "dark gas" (interpreted as H$_2$ without CO) for the same regions. We find that the cloud masses deduced from millimeter dust emission are fully compatible with the total masses of the clouds (the "dark gas" plus molecular gas traced by CO emission). However, for the regions chosen in this study, the fraction of "dark gas" with respect to the total molecular gas is low. 

\citet{SRB+87} compared, for a large sample of giant molecular clouds in our Galaxy, the virial mass of the clouds to their CO luminosity. They observed a good correlation between both quantities that can be transformed into a relation between the mass deduced from CO emission and the virial mass. Since in the studied regions the masses deduced from CO emission are close to the masses deduced from millimeter dust emission, this law can then be compared to our mass estimates for the Galactic clouds  (see Fig. \ref{fig3}).  For the Galactic clouds, the millimeter and CO masses are consistent for the Galactic gamma ray calibration of  $X_{CO}=1.8\cdot 10^{20}$ mol.cm$^{-2}$.(K.km.s$^{-1}$)$^{-1}$. For this factor, the \citet{SRB+87} virial mass-L$_{CO}$ relation show that the virial mass is larger than the cloud mass by a factor decreasing toward high masses (see Fig. \ref{fig3}). This trend was also observed for  clumps within GMCs by numerous studies in our Galaxy \citep{BM92,Williams:1995lr,Heyer:2001fk,Simon:2001lr}. In the Galaxy, the $f$ factor is observed to be 1 only for the most massive Galactic GMCs with masses larger than $5\cdot 10^5 M_\odot$ \citep{Dame:1986uq}. For clumps within large gravitationally bound clouds the weight of the overlying material in the cloud could provide significant binding pressure for the clumps \citep{BM92}.

\subsection{In the SMC}

The comparison of gas masses obtained from the millimeter dust emission with the virial masses shows a systematic bias in the SMC. Gas masses deduced from the millimeter emission are 1.8 times higher on average than the virial masses. 
The factor $f=\frac{\mathrm{M_{1.2mm}}}{\mathrm{M_{vir}}}$ ranges from 0.9 to 3.1 (see Table \ref{tbl-4}). The  mass discrepancy reported for SMCB1\#1 by \citet{RBR+04} is thus a general result that must be understood. 

For the computation of masses from the millimeter flux, we used a dust temperature of 15~K for all the clouds. Could a higher temperature in the molecular cloud of the SMC explain the difference observed in mass? For agreement between the virial mass and the mass from millimeter dust emission to agree, the dust temperature within the molecular gas would have to be 23~K on average. But if the molecular clouds of the SMC were warm, they would emit in the far infrared. While for most of the regions surveyed here, the Spitzer 160$\mu$m emission \citep{Leroy:2006fk} is biased toward the H{\sc ii} region emission present in the beam, this is not the case for the quiescent molecular cloud SMCB1\#1. Comparing millimeter fluxes with Spitzer data in SMCB1\#1 gives a dust temperature of 11, 14 or 17~K for spectral indices of 2, 1.5 and 1, respectively. This shows that at least in a quiescent cloud (and presumably in all molecular clouds) a higher dust temperature in the molecular clouds of the SMC is not compatible with the far-infrared data and can not explain the difference observed between the mass estimates.

Furthermore, the cloud masses deduced from dust millimeter emission depend on the dust-to-gas ratio and the emissivity of dust grains in the SMC molecular clouds. The values we used for these parameters  were chosen to limit the mass of the clouds and thus reduce the discrepancy with virial masses. In the diffuse ISM of the SMC, the dust emissivity per hydrogen atom is measured to be 1/30 the emissivity of the  Galactic diffuse medium \citep{BBL+04}. For the molecular clouds of the SMC, we used a larger emissivity to account both for grain coagulation effects (twice the diffuse medium value) and for our estimate of the SMC dust-to-gas ratio (twice larger than the canonical value for the metallicity of the SMC, c.f. Appendix \ref{append2}).  Not only it is not clear if grain coagulation (grain emissivity enhancement by mass unit) occurs in the SMC, but we assume that heavy elements depletions are the same in the SMC and Galactic molecular clouds while they are observed to be smaller in the SMC diffuse ISM. In that sense, we consider the millimeter mass estimates in Table \ref{tbl-4} to be lower limits on the molecular cloud masses. Lower depletions and thereby lower dust emissivity per hydrogen atom will enlarge the mass discrepancy. The range of masses obtained for each cloud, for dust emissivities ranging from 1/6 to 1/20 the Galactic molecular value, is shown in Fig. \ref{fig3}. Since it is dynamically difficult to account for a large difference between the cloud mass and its virial estimation, a first conclusion of this paper is that SMC dust observations imply a significant (up to a factor $\sim 5$) increase in the dust-to-gas mass ratio from the diffuse ISM to molecular clouds. The dust which is destroyed by shocks in the diffuse ISM is rebuilt in molecular clouds. A second conclusion is that even with conservative assumptions on the dust emissivity per hydrogen atom, the virial mass is found to be systematically lower than the millimeter gas mass. 

\section{Generalized virial theorem}

In the SMC, molecular cloud masses from millimeter dust emission are systematically larger than the virial masses. We saw above that uncertainties on the mass estimates from dust emission can not explain such differences. Observational estimates of the velocity dispersion are obtained through the CO line width. The lack of resolution and sensitivity of the observations certainly plays a role on the virial mass computations. However, the observational bias on the Galactic and the SMC sample should be similar and we were careful to apply the same analysis, so we are confident that  the virial mass discrepancy in the SMC does not arise from observational uncertainties. 

\citet{Leroy:2006fk} studied dust emission in the far-infrared in the SMC and showed that excesses with respect to H{\sc i} are observed to extend farther than CO emission observed with NANTEN. This difference is attributed to molecular hydrogen not traced by CO and amounts to 30\% in spatial coverage. The CO and SIMBA observations we use for the SMC clouds certainly underestimate the size of the giant molecular clouds due to sensitivity effects. As a result, the masses measured in this study are characteristics of clumps within the clouds smaller than the one from the \citet{Leroy:2006fk} study. However, the size of the clouds that are detected with SIMBA and CO observations (see Fig. \ref{fig1}) in our study are about the same, so that the effect of any biases in the cloud radius should not be large enough to affect our results.

 The mass discrepancy observed between the SMC and Galactic clouds might indicate that the CO line width underestimates the amplitude of the cloud turbulent motions. This interpretation was proposed by \citet{RBR+04}. In the SMC clouds, the CO emission is expected to come from dense clumps embedded in lower density gas where CO is photo-dissociated \citep{LLD+94}. The bulk of the cloud mass is in the lower density gas but the CO line width only measures the velocity dispersion of the clumps. The same statement applies locally to the less massive high Galactic latitude clouds where CO line widths are observed to be systematically lower than the H{\sc i} line width. Quantitatively, in the Ursa Major high latitude cloud, on scales of 5-10 pc, the H{\sc i} velocity dispersion is observed to be twice that of CO \citep{de-Vries:1987lr}. It is not clear to us whether this observed difference can be simply extrapolated to the more massive SMC clouds that are gravitationally bound. However, it has been theoretically suggested that in gravitationnal bound clouds the gas velocity dispersion should increase as the density decreases \citep{Fatuzzo:1993fk,McKee:1999qy}. In this section, we present an alternative interpretation where the mass difference results from the contribution of the magnetic field to the clouds support.

In this section, we inspect  the assumptions made for the virial mass calculations, searching for those which could cause a systematic underestimation of the mass. In particular, the presence of a magnetic field in the molecular clouds could help to balance gravity and we will compute the necessary strength to explain the mass ratio $f$ observed in the SMC.

To apply the virial theorem, molecular clouds should be virialized. Observationally determined dynamical parameters enable us to define the characteristic time scale of these clouds: $R/\Delta V$. Clouds younger than this characteristic time didn't have time to be virialized and the virial theorem can not apply. This characteristic time scale for the SMC clouds is $\sim 3\cdot 10^6$ years. It is therefore possible that these clouds did not have time to be virialized. One possibility would be that, unlike Galactic giant molecular clouds, in the SMC CO observations only traces clouds collapsing into stars on the timescale of a few million years (free-fall time). But this interpretation does not fit with the fact that two clouds, SMCB1\#1 and Hodge15, have no associated H{\sc ii}.

In this study, virial masses were determined using the simplified form from \citet{MRW88}. This form neglects among others the surface pressure and the magnetic field terms. 
The general virial equilibrium equation for a cloud in a steady state can be written as:
\begin{equation}
0=2(\mathcal{T}-\mathcal{T}_0)+\mathcal{M}+\mathcal{W}
\label{eq:viriel}\end{equation}
\citep{MZ92} where $\mathcal{T}$ is the total kinetic energy (due to thermal and global motions), $\mathcal{T}_0$ represents the confinement by external thermal pressure, $\mathcal{M}$ is the magnetic energy and $\mathcal{W}$  is the gravitational energy.
The gravitational energy can be expressed as:
\begin{equation}
\mathcal{W}=-\frac{3}{5}a \frac{GM^2}{R}
\end{equation}
 where G is the gravitationnal constant and $a$  measure departures from uniform density and sphericity \citep{BM92}. 

\citet{MZ92} introduced the virial parameter $\alpha=a\frac{2\mathcal{T}}{| \mathcal{W}|}$ so that the mass of a cloud is expressed as:
\begin{equation}
M=\frac{5\sigma^2 R}{\alpha G}.
\end{equation}

 If we take $M_{1.2mm}$ as the true gas mass of the cloud, understanding the high values of $f=M_{1.2mm}/M_{vir}$ amounts to looking for low values of the virial parameters (with $f=10/9\alpha$), which is equivalent to solving:
\begin{equation}
f=\frac{10}{9 a}\frac{\left[1-(P_0/\overline{P})\right]}{\left[1-(\mathcal{M}/|\mathcal{W}|)\right]}.
\label{eq:fparam}\end{equation}

The values of f are then affected by density and geometry effects ($a$), external pressure ($P_0$) and magnetic fields ($\mathcal{M}/|\mathcal{W}|$).  As we saw in Galactic giant molecular clouds, $f$ is smaller than unity and is interpreted as external pressure confinement. Since external pressure (high $P_0$)  acts in lowering $f$ and the effects of density and geometry are mild, we will only discuss the effects of  magnetic field support in the SMC giant molecular clouds. 

\subsection{Magnetic field support in clouds. \label{sec:magfield}}

From the observed characteristics of the molecular clouds, we can compute the magnetic-to-gravitational energy ratio needed to explain the $f$ factors. Taking the case of a spherical uniform cloud with steady motions and a mean pressure exceeding the surface pressure ($\bar{P}>>P_0$), one obtains  $\mathcal{M}/|\mathcal{W}|=1-\frac{10}{9f}$. The obtained values are displayed in Table \ref{tbl-5} .

\begin{table}
\centering
\caption{Deduced characteristics for the SMC clouds. \label{tbl-5}}
\begin{tabular}{llllll}
name &  $f$ & $\mathcal{M}/|\mathcal{W}|$ & $|B|$ & $n_H$\\
 & &  & $\mu$G & $10^3$cm$^{-3}$\\
\hline
LIRS 49& 2.2 & 0.5 & 24 & 0.8\\
LIRS 36& 1.4 & 0.2 & 9 & 0.3\\
SMCB1\#1& 2.0 &0.5 & 12 & 0.5\\
SMCB1\#2 & 1.8 & 0.4 & 11 & 0.3\\
Hodge 15& 0.9 & -- & -- & 0.4\\
SMCB2 S & 3.1 & 0.6 & 20 & 0.3\\
SMCB2 N & 1.5 & 0.2 & 11 & 0.4\\
\end{tabular}
\end{table}

Taking $\mathcal{M}=\frac{1}{6}(B^2-B_0^2)R^3$, one can express the strength of the magnetic field that is necessary to account for the $f$ values:
\begin{equation}
B^2-B_0^2=\frac{18}{5}\left(1-\frac{10}{9f}\right)\frac{GM^2}{R^4}
\end{equation}
where $B_0$ is the ambient medium value of the magnetic field strength. Taking $B_0=5\mu$G $^($\footnote{consistent with \citet{Poh93} model; note that this choice does not really influence the result as long as the mean SMC magnetic field remains low, which we have no reason to doubt}$^)$, the magnetic field strength necessary to support the clouds is computed and displayed in Table \ref{tbl-5}. They range from $\sim 10$ to $\sim30~ \mu$G, with a mean value of $\approx 15~ \mu$G. Such strengths are similar to the values reported for some Galactic clouds \citep{MG88a} but the clumps surveyed are much smaller and confined by the pressure, not by their self-gravity \citep{BM92}. Recently,  a magnetic field  with a similar strength ($\sim 20\mu$G) has been observed in the outer part of the Taurus molecular cloud \citep{Wolleben:2004lr}.

So  magnetic field support of the molecular clouds could explain the mass difference observed in the SMC. But why would the magnetic field support be important for the SMC molecular clouds and not for the Galactic ones?  Molecular cloud formation could be different in the SMC. Giant molecular clouds in spiral galaxies are thought to form in spiral arms where the low shear and high gas density allows gravity to gather diffuse matter over large scales. Matter may preferentially condense along field lines without compressing the magnetic field. We refer to \citet{Rosolowsky:2003uq} for a discussion of observational constraints on proposed formation scenarios based on a survey of molecular clouds in M33. In an irregular galaxy like the SMC, massive stars have an important impact on the structure of the interstellar medium and molecular clouds may form out of shells blown by stellar winds and supernovae explosions. In the Spitzer infrared images \citep{Bolatto:2006kx}, the molecular clouds in Fig. \ref{fig1} do appear as local peaks along a large shell of 300pc diameter. In this picture, the magnetic field is expected to be swept together with the matter.  It is also possible that the lower metallicity and therefore the lower dust content in the SMC creates lower dust opacity in the GMCs of the SMC than in our Galaxy. The ionization could therefore be higher in the SMC, supporting higher magnetic fields. Another possibility is that magnetic field support is important also in the Galactic molecular clouds and virial masses could have to be re-examined taking into account a magnetic field support on large scales. In this case, it would be interesting to see if the larger external pressure that will be required for the virial mass to match the other mass estimates (from $\gamma$ ray analysis for example) is compatible with observations.  Answering these questions are beyond the scope of this paper but will can be addressed in the future. In any case, magnetic field observations (e.g through Zeeman effect) can be used to confirm the magnetic field strength in these regions.

\section{Conclusions}

We compared two different molecular gas tracers, the millimeter dust emission and the virial theorem applied to CO observations, at two different metallicities: our Galaxy and the SMC. SIMBA/SEST millimeter observations for the SMC and FIRAS observations for the Galaxy have a similar linear resolution, enabling a uniform study on similar giant molecular clouds in these two environments.

We find a clear difference between the two samples: in the Galaxy, the virial mass is systematically larger than the mass deduced from millimeter emission, while in the SMC sample, the virial mass is lower than the mass deduced from millimeter emission. The main uncertainty in the mass determination comes from the dust emissivity per hydrogen atom in the SMC. We have used the millimeter emissivity determined in Galactic molecular gas scaled by 1/6 to account for the lower SMC metallicity. This scaling factor has been calculated asssuming similar depletions of heavy elements in the SMC and in the Galaxy. We consider this assumption to be conservative because diffuse ISM depletions have been inferred to be smaller in the SMC than in the Galaxy from UV spectroscopy and dust observations. The millimeter cloud masses would have been larger for smaller depletions (i.e. smaller dust-to-gas ratio), enhancing the discrepancy.

For Galactic clouds the dust and CO masses are consistent with the gamma ray detemination of the Galactic $X_{CO}$ factor. The virial masses are higher than the cloud masses. The underestimation of SMC molecular cloud masses by virial methods is thus an unexpected result. Inspecting the assumptions made in the virial mass computations, we find that magnetic field support in the clouds could explain the difference between the two mass estimates. In this scenario, SMC giant molecular clouds are not fully supported by turbulent motions. Additional support is given by a magnetic field of $\sim 15\mu$G. Further studies and magnetic field observations are needed to confirm this interpretation.

An alternative explanation is that the virial mass is underestimated because the CO line width does not measure the full amplitude of the turbulent motions within the cloud. In the SMC GMCs, the CO emission is expected to come from dense clumps embedded in lower density gas where CO is photo-dissociated. The bulk of the cloud mass is in the lower density gas, but the CO line width only measures the velocity dispersion of the clumps. Can the lower density gas turbulence be higher than the clump to clump velocity dispersion? 
This study emphasizes the power of millimeter emission as a dense matter tracer. The large amount of millimeter data coming will enable interesting follow up studies with better resolution and sensitivity.

\begin{acknowledgements}
The author would like to thank T. Dame for providing us with the CO Galactic survey data cube and I. Grenier for helping with the comparison with the gamma ray results.
This study has been supported but the french national program PCMI (Physique et Chimie du Milieu Interstellaire, CNRS). M.R. is supported by the Chilean Center for Atrophysics FONDAP No. 15010003. C.B. and F.B. wish to aknowledge the Chilean Center for Astrophysics FONDAP No 150010003 for support of their visit in Chile.
\end{acknowledgements}

\bibliographystyle{aa}
\bibliography{./aamnem99,./biblio}

\begin{appendix}

\section{Emissivity of dust associated with molecular gas}\label{append1}

Dust emissivity per hydrogen atom is often taken in the diffuse medium where it is easy to deduce directly from observations. However, dust properties could vary from diffuse to dense regions \citep{CBL+01,SAB+03,Cambresy:2005vl}. We therefore chose to recompute the emissivity of dust is computed in the molecular ring in order to represent the properties of dust associated with molecular hydrogen. 

In the molecular ring, the 1.2mm dust continuum and CO emissions longitude profiles along the Galactic plane are well correlated and the ratio of their intensities is constant: $R=(0.89\pm 0.01)\cdot 10^{-1} MJy(K.km.s^{-1}.sr)^{-1}$ (Boulanger et al., in prep.). This ratio is related to the well known CO-to-H$_2$ conversion factor ($X=N(H_2)/I_{CO}$) through the following equation:
\begin{equation}
R=2\epsilon_H(\lambda)B_\lambda(T_{dust}) X,
\end{equation}
where $\epsilon_H$ is the dust emissivity per hydrogen atom in the observed region. Although $T_{dust}$ may vary from one cloud to the other, these variations are likely to be moderate and can therefore be neglected at millimeter wavelenght while $T>10K$. In this study, we consider the temperature range measured by FIRAS as pertinent for molecular clouds analysis.
The CO-to-H$_2$ factor is empirically determined in different environments and is largely discussed in the litterature.

In our Galaxy and in particular in the solar neighbourhood, studies of $\gamma$ rays and gas tracers \citep{SBD+88} enable a determination of this $X$ factor. For our Galaxy as a whole, the analysis of recent observations gives $X$ factors between $1.56$ and $1.9\times 10^{20} mol.cm^{-2}.(K.km.s^{-1})^{-1}$ \citep{SM96,HBC+97,GCT05}. We use the recent estimate obtained for the local molecular clouds  through $\gamma$ rays analysis (Grenier et al., priv. comm.): $X=1.8 \pm0.3 \cdot 10^{20} mol.cm^{-2} (K.km.s^{-1})^{-1}$.

Using this local Galactic X factor and using the ratio of millimeter and CO intensities determined in the molecular ring, we compute the emissivity of dust associated with molecular gas in our Galaxy:
\begin{equation}
\epsilon_H^{ref}(1.2mm,H_2)=(1.20\pm 0.20)\times 10^{-26} at^{-1}.cm^{2}
\end{equation}
This value is used in eq. \ref{eq:emimm} to compute molecular cloud masses.

The dust emissivity in the diffuse medium is measured through the correlation at high Galactic latitude betwwen dust emission and H {\sc i} emission at low column densities \citep{BAB+96}. For a dust temperature of 17.2K, this correlation gives a dust emissivity per hydrogen atom of:
\begin{equation}
\epsilon_H^{ref}(1.2mm,HI)=(6.7\pm 1.2)\cdot 10^{-27} at^{-1}.cm^2.
\end{equation}

Dust emissivity in the molecular medium is therefore twice larger than in the diffuse medium. Such an increase in emissivity from the diffuse medium to molecular clouds was observed in our Galaxy, but on much smaller  and denser regions \citep{CBL+01,SAB+03}.  This increase is observed to be correlated with a drop small grain abundance. It has been thus interpreted as the coagulation of small grains between themself or on large grains in dense regions of molecular clouds, giving composite porous grains with a larger emissivity.

\end{appendix}
\begin{appendix}

\section{The dust to gas ratio in the SMC}\label{append2}

The gas-to-dust ratio in the SMC has been determined by using extinction curves to derive N(H {\sc i})/A(V) values \citep{GCM+03,BLM+85,TSR+02}. However, the derived values correspond to specific line of sights (where extinction can be measured), they rely on the present extinction law understanding, and there is a significant scatter from one region to the other and from one study to the other \footnote{The dust to has ratio found this way to range between $\sim 1/6$ and $\sim 1/30$ the Galactic value}. \citet{BBL+04} deduced a gas-to-dust ratio 30 times the solar value by comparing the dust and H {\sc i} emission of the SMC. However, this value refers to the diffuse medium and could reflect dust grain destruction processes that do not apply so intensively in the denser environnements of the SMC bar where grains are more shielded.
 A generally used value for the SMC dust-to-gas ratio is one tenth of the solar one, using the metallicity difference \citep{Duf84}. However, it relates to gas phase abundances while we are interested in dust masses and most importantly since 1984 the solar abundances have been updated \citep{SM01a,SM01b}. 

We therefore choose to recompute the SMC dust abundances to evaluate the dust-to-gas mass ratio difference between the SMC and the Solar Neighbourhood. For that, we assume that the elemental depletions are the same in the SMC as in the Solar Neighbourhood. This is supported by studies of Sk 108 in the main bar of the SMC, where the depletion patterns are similar to those of Galactic halo clouds \citet{WLB97,MWY+01}. The depletions in the SMC could be less than the solar values \citep{RVT+03}, so that our estimation could be an upper limit. Comparing the $\zeta$ Oph dust and solar abundances given by \citet{SM01a,SM01b}, we inferred depletions $\eta$ of 0.66, 0.11, 0.46, 0.92, and 0.95 for C, N, O, Mg, and Si respectively.

The main sequence star AV 304 shows no sign of processing in its photosphere and could therefore represent the present-day total (gas and dust) composition of the SMC \citep{RVT+03}. Using \citet{TLP+04} corrected abundances for this star with the depletion factors computed above, the dust-to-hydrogen mass ratio can be computed:
\begin{equation}
\frac{m_{dust}}{m_{H}}=\sum_{M} [M/H]_{tot}\cdot \eta_M\cdot \frac{m_M}{m_H},
\end{equation}
 and is used as an estimate of the dust-to-gas mass ratio.
We obtained a dust-to-hydrogen mass ratio of $1.51 \cdot 10^{-3}$ for the SMC. Comparing this value to the dust-to-hydrogen mass ratio for $\zeta$ Oph dust\footnote{$m_{dust}/m_{H}(\zeta Oph)=8.88\cdot 10^{-3}$ as computed from \citet{SM01a,SM01b} abundance values.}, the dust-to-gas mass ratio difference between the SMC and the Solar Neighbourhood is found to be:
\begin{equation}
\frac{x_{SMC}}{x_{\odot}}=0.17
\end{equation}

\end{appendix}

\end{document}